\documentclass[twocolumn,floatfix,rmp]{revtex4}

\usepackage{graphics}
\usepackage{epsfig}
\usepackage{amsmath}
\usepackage{color}

\begin{document}

\title{Under the influence of alcohol: The effect of ethanol and methanol on 
lipid bilayers}

\author{Michael Patra}
\affiliation{Biophysics and Statistical Mechanics Group,
Laboratory of Computational Engineering, Helsinki University
of Technology, P.\,O. Box 9203, FIN--02015 HUT, Finland}

\author{Emppu Salonen}
\author{Emma Terama}
\author{Ilpo Vattulainen}
\affiliation{Laboratory of Physics~and~Helsinki Institute of Physics,
Helsinki University of Technology, P.\,O. Box 1100, FIN--02015 HUT, Finland}

\author{Roland Faller}
\author{Bryan W. Lee}
\affiliation{Department of Chemical Engineering and Materials Science,
University of California-Davis, One Shields Ave, Davis, CA 95616, USA }

\author{Juha Holopainen}
\affiliation{Department of Ophthalmology, University of Helsinki, Finland and
Helsinki Biophysics \& Biomembrane Group, Institute of Biomedicine, University of Helsinki, Finland}

\author{Mikko Karttunen}
\affiliation{Biophysics and Statistical Mechanics Group,
Laboratory of Computational Engineering, Helsinki University
of Technology, P.\,O. Box 9203, FIN--02015 HUT, Finland }

\begin{abstract}

Extensive microscopic molecular dynamics simulations have been performed to
study the effects of short-chain alcohols, methanol and ethanol, on two
different fully hydrated lipid bilayer systems in the fluid phase at
$323~\mathrm{K}$. It is found that ethanol has a stronger effect on the
structural properties of the membranes. In particular, the bilayers become more
fluid and permeable: Ethanol molecules are able to penetrate through the
membrane in typical time scales of about $200~\mathrm{ns}$ whereas for methanol
that time scale is considerably longer, at least of the order of microseconds. 
We find good agreement with NMR and micropipette studies. We have also
measured partitioning coefficients and the rate of crossing events for
alcohols, i.\,e., typical time scale it takes for a molecule to cross the lipid
bilayer and to move from one leaflet to the other. For structural properties,
two-dimensional centre of mass radial-distribution functions indicate the
possibility for quasi long-range order for ethanol--ethanol correlations in
contrast to liquid-like behaviour for all other combinations.

\end{abstract}

\maketitle

\section{Introduction}
\label{secIntroduction}

It is well known that even small changes in the composition of cell
membranes can strongly affect the functioning of intrinsic membrane
proteins, such as ion and water channels, which regulate
the chemical and physical balance in cells~\cite{mazzeo88,cantor03}.
Such changes may occur due to the introduction of
short-chain alcohols, or other anaesthetics, at membrane surfaces.
Although anaesthetics are being used every single day in hospitals
around the world, the molecular level mechanisms of general
anaesthesia remain elusive, 
see e.\,g.~\cite{cantor97,eckenhoff01,tang:2002}. The same applies to the
effect of alcohols on biological systems. \citet{Klemm:1998}
provides a good review of the topic.

Another aspect to the effect of alcohols appears in a more
applied context. In the process of producing alcoholic beverages, wine in
particular, yeasts like \textit{saccharomyces cervisiae} have to
sustain high ethanol concentrations without losing their
viability. However, in about $10\,\%$ of all wine fermentations 
the industry encounters so-called stuck
fermentations~\cite{bisson02,Silveira:2003}. There is no
satisfactory understanding of this effect. Some
models propose that an effect very similar to general anaesthesia
is responsible for rendering the yeast
cells dormant~\cite{cramer02}. It has been suggested that high 
alcohol concentrations 
change the membrane structure and force transmembrane
proteins into unfavourable conformations. In these conformations
proteins cannot fulfil their functions and thus the yield drops
dramatically.

In addition to the above aspects, there are other important issues as well. In
particular, in cellular systems such as bacteria and yeast, the toxicity of
ethanol has been suggested to be due to its interaction with
membranes~\cite{ly02,Silveira:2003,ly-pre} and the consequent general effects
such as changes in mechanical properties, permeability and diffusion. Changes
in such generic membrane properties may affect the functions of proteins and
binding sites due to changes in lateral pressure~\cite{eckenhoff01}, or, if the
membrane becomes more permeable, changes in the electrostatic potential may
affect signalling. These effects are not to be mixed up with the toxicity due
to metabolic products such as acetaldehyde from consumption of ethanol -- the
cause of poisoning commonly known as hangover.

We concentrate on the effects of ethanol and methanol on
structural properties of membranes. It is quite surprising that
despite a vast number of clinical and biochemical studies, there
has been very few computational investigations of the effect of
short-chain alcohols, or other anaesthetics, on membranes. The
only simulational studies of bilayers and ethanol are, 
to the authors' knowledge,
the one by \citet{Feller:2002} who
used molecular dynamics simulations of ethanol and POPC
(palmitoyloleoylphosphatidylcholine) lipid bilayers and NMR to
study the molecular level interactions in these systems, and
the article by \citet{bwlee04} discussing 
alcohol--membrane systems briefly.
Direct comparison of our results with Feller~et~al.\ 
is not meaningful since
their study was performed using a different ensemble close to the
gel state at $283~\mathrm{K}$ whereas
here we are in the biologically relevant fluid phase at $323~\mathrm{K}$.
For methanol--bilayer systems there exists to the authors' knowledge
only one computational
article~\cite{Bemporad:2004}.

For anaesthetics the situation is slightly better. 
\citet{tang:2002}
used molecular dynamics simulations to study molecular level
mechanisms of general anaesthesia using halothane as a specific
anaesthetic. They concluded that the global effects of
anaesthetics, i.\,e., due to generic interaction mechanisms, 
are important and lead to modulations in the functions of channels
and\,/\,or proteins. These conclusions are also supported by the fact
that the same anaesthetics are effective for humans
and a variety of animals. Similar conclusions for halothane interactions
with bilayers have been 
pointed out by \citet{koubi00,koubi01}.
The importance of generic effects has also
been indicated in recent experimental studies of the effect of
ethanol on \textit{Oenococcus oeni} cells~\cite{Silveira:2003}.
Although the shortage of simulational studies may be due to the
high computational demands of these systems, it is still
surprising since computer simulations can provide detailed
information about fundamental molecular level mechanisms.

In this article we study the effect of two short-chain alcohols,
ethanol and methanol,
on two different lipid membranes consisting of either pure DPPC
(dipalmitoylphosphatidylcholine) or POPC.
Methanol is a small solute having a single
hydrophilic hydroxyl group whereas ethanol possesses an additional hydrophobic
carboxyl group. DPPC and POPC share the same headgroup 
but one of the tails of POPC
has a double bond and is two carbon atoms longer,
whereas DPPC has only single bonds in its chains, see Fig.~\ref{figChemie}.
We have studied these systems under fully hydrated conditions using
microscopic molecular dynamics. $50~\mathrm{ns}$ trajectories for each of the
four combinations of lipid and alcohol allow us to gather
high statistical accuracy.

Phospholipid bilayers can be considered as a first approximation to understand
the behaviour of cell membranes under the influence of alcohol, and much
information can be extracted from such systems. The simulations show that
ethanol is able to pass through the bilayer much more easily than methanol.
This can be explained by the hydrophobic nature of the carbon ``tail'' of
ethanol, making passing through the hydrophobic tail regions of lipid bilayers
easier. In addition, ethanol molecules condense near the interface region
between lipids and the surrounding water, i.\,e., there is a sharply increased
density of ethanol near the interface region, while for methanol only a
moderate increase of the density is seen near the interface region. These
effects are very pronounced for DPPC bilayers, and only slightly weaker for
POPC bilayers. This has far reaching implications for the basic properties of
bilayers.

The rest of this article is organised as follows. In the next section we 
describe the model and the simulation details. Then, in Sec.~\ref{secResults},
we present the results from the simulations. Section~\ref{secDiscussion}
contains a discussion and conclusions.

\section{Model and simulation details}
\label{secModel}

{\small

We have simulated lipid bilayer systems consisting of either 128 DPPC or 128
POPC molecules (i.\,e., 64 lipids in each leaflet). For the lipids we used a
previously validated united atom model~\cite{tieleman:96a}. The DPPC
simulations are based on the final structure of a $100~\mathrm{ns}$ run of a
DPPC bilayer that is fully hydrated by 3655 water molecules. The configuration
is available online\footnote{http://www.softsimu.org/downloads.shtml}. The
simulations by \citet{patra:03aa} were run using the same parameters as here
(details below), and the $100~\mathrm{ns}$ run used in this study is a
continuation of a $50~\mathrm{ns}$ study~\cite{patra:03aa,falck:04a}. For the POPC
simulations, such an initial structure had to be generated first. We started
with a fully hydrated POPC bilayer~\cite{tieleman:99a} and simulated it for
$10~\mathrm{ns}$. The final structure of that simulation run was used as a
starting point for the POPC simulations reported here.

In order to add the alcohol molecules, the simulation box was first extended in
$z$-direction such that an empty volume was created. A total of $90$ ethanol
(methanol) molecules were randomly inserted in the empty volume, and the
remaining space was filled with water. The total number of water molecules then
amounted to 8958 for the DPPC systems and 8948 for the POPC systems, or, in
other words, 1 mol\% alcohol on a lipid free basis. The small difference
between DPPC and POPC systems is due to the different lateral extensions of the
bilayers.

The force field parameters for bonded and non-bonded interaction
were taken from \citet{berger:97a}, available
online\footnote{http://moose.bio.ucalgary.ca/Downloads/files/lipid.itp}.
Partial charges were taken from~\citet{tieleman:96a},
available online for both DPPC~\footnote{http://moose.bio.ucalgary.ca/Downloads/files/dppc.itp}
and POPC~\footnote{http://moose.bio.ucalgary.ca/Downloads/files/popc.itp}.
As is seen in the chemical structures in Fig.~\ref{figChemie},
DPPC and POPC are identical up to a single pair of CH-groups,
connected by a double bond instead of a single bond in the \textit{sn}\,--\,2 chain of POPC,
and the two additional CH$_2$ groups at the end of that chain. This
similarity is reflected in the force fields, which are identical
up to the modelling of the four affected atoms. Ethanol and methanol were modelled using the Gromacs force
field parameters~\cite{lindahl:01a} which are identical with the exception of
the added CH$_2$ group for ethanol. 
Thus, differences
observed between the two lipids or the two alcohols do not originate from
differences in their respective force field parameterisations but
are due to the physics and\,/\,or chemistry of those components.
For water the Simple Point Charge (SPC) model~\cite{berendsen:81a} was
used.

\begin{figure}
\centering
\includegraphics[width=3.2in]{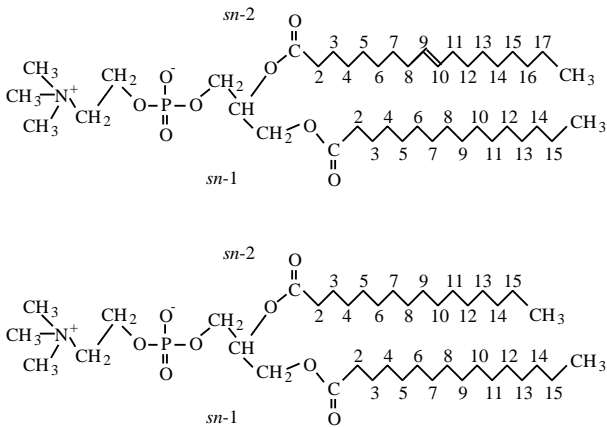}
\caption{Structures of POPC (top) and DPPC (bottom). They are identical with
the exception of the \textit{sn}\,--\,2 chain which is two carbons longer and 
contains one double bond for POPC.}
\label{figChemie}
\end{figure}

The simulations were performed with the Gromacs package~\cite{lindahl:01a}. The
lipids, water molecules and alcohols were separately coupled to a heat bath at
temperature $T=323~\mathrm{K}$ using the Berendsen
thermostat~\cite{berendsen:84a} with a coupling time constant of
$0.1~\mathrm{ps}$. All the bond lengths were constrained to their equilibrium
values by the Lincs algorithm~\cite{hess:97a}. Pressure was controlled using
the Berendsen barostat~\cite{berendsen:84a} with a time constant of
$1~\mathrm{ps}$. 
The pressure coupling was used semi-isotropically such that height of the
box ($z$ direction) and the cross sectional area ($x y$-plane) were allowed
to vary independently of each other. 
Lennard-Jones interactions were cut off at a distance of
$1.0~\mathrm{nm}$ and the time step was set to $2~\mathrm{fs}$. Long-range
electrostatics were updated every 10-th time step [the twin-range
scheme~\cite{kessel,patra:03b} was used], and handled by the particle-mesh
Ewald (PME) algorithm~\cite{essman:95a}. For DPPC bilayers it has been shown
that replacing PME by the computationally cheaper cut-off scheme leads to
pronounced artifacts~\cite{patra:03b,patra:03aa}.

The systems were simulated for a total of $50~\mathrm{ns}$ each. After
$20~\mathrm{ns}$, the samples had equilibrated,
and the remaining $30~\mathrm{ns}$ were used for data collection.
Equilibration was determined by monitoring the area per lipid
as described in the next section. 
For completeness, we also present
results for pure DPPC and POPC bilayers. In particular the
latter ones are important since many of the results have so far only been
published based on simulations using a cutoff for handling
electrostatics. In addition to the above systems, we also performed
a control simulation with a dehydrated system containing only 10 water
molecules per lipid. This was done in order to see if dehydration
has a direct effect on the properties but no significant effects were found.

The simulations took a total of 16~000 CPU hours using an
IBM \mbox{eServer} Cluster 1600 (Power4 processors).

}

\section{Simulation results}
\label{secResults}

Before presenting a systematic summary of our results, we 
give a quick overview of the basic properties of these systems.
Alcohol molecules have a tendency to collect in or near the bilayer
(Sec.~\ref{secMassDensity}). This tendency is stronger for ethanol than 
for methanol as confirmed by a partition analysis
(Sec.~\ref{secPartitioning}).
Ethanol is able to form hydrogen bonds with the lipids in 
the bilayer (Sec.~\ref{secBinding}), and these hydrogen bonds reduce the order 
parameter of the lipid hydrocarbon tails. The combination of all this
results in an easy penetration of
ethanol through the bilayer. In contrast, no hydrogen bonds or penetration
was observed for methanol.

In this paper we use the following colour code for all figures. Curves for
systems containing {\color{red}ethanol are drawn in red}, curves for
{\color{green}methanol in green} and pure lipid systems
{\color{blue}without alcohol in blue}.

\subsection{System dimensions}
\label{secDimensions}

\begin{figure}[b]
\centering
\includegraphics[height=3.8cm]{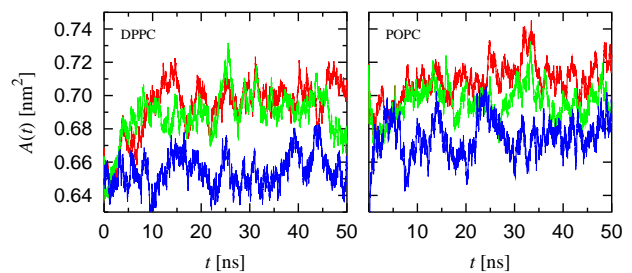}
\caption{Temporal behaviour of the area per lipid $A(t)$
for a DPPC bilayer (left) and a POPC bilayer (right).
The colour of the line marks whether the lipid bilayer has been simulated in the
presence of ethanol (red), of methanol (green), or of no alcohol (blue).
This is our standard colour code employed throughout this paper.}
\label{figAreaPerLipid}
\end{figure}

The area per lipid is one of the most important quantities characterising
lipid bilayer systems and it can also be used to monitor equilibration
during a simulation run. 
The time evolutions
of the area per lipid in the systems studied here are shown in
Fig.~\ref{figAreaPerLipid}. The average areas per lipid, $\langle A \rangle$, obtained in our
simulations are listed in Table~\ref{tabArea}.

\begin{table}
\begin{tabular}{l|c}
\hline \hline
System   & Average area per lipid \\
\hline
DPPC (pure)  & $(0.655 \pm 0.002) ~\mathrm{nm}^2$ \\
DPPC + ethanol & $(0.699 \pm 0.002) ~\mathrm{nm}^2$ \\
DPPC + methanol & $(0.693 \pm 0.004) ~\mathrm{nm}^2$ \\ \hline
POPC (pure)  & $(0.677 \pm 0.003) ~\mathrm{nm}^2$ \\
POPC + ethanol & $(0.699 \pm 0.003) ~\mathrm{nm}^2$ \\
POPC + methanol & $(0.693 \pm 0.003) ~\mathrm{nm}^2$ \\
\hline \hline
\end{tabular}
\caption{
Average area per lipid for all systems studied in this work. A weak
effect of the alcohols is visible. The error estimates have been computed
from block averaging and extrapolating to large block sizes.} 
\label{tabArea}
\end{table}

For a pure DPPC bilayer we obtain
$\langle A_\mathrm{DPPC}\rangle = 0.655~\mathrm{nm}^2$
agreeing well with previous simulations and experiments, see
Ref.~\cite{patra:03b} and references therein. For pure POPC
we obtain $\langle A_\mathrm{POPC}\rangle = 0.677~\mathrm{nm}^2$
in agreement with previous computational studies~\cite{chiu99,pasen03}
and slightly larger than the results from x-ray diffraction
studies~\cite{pabst:00a,pabst:00b}.
For POPC, the difference to x-ray diffraction results
$\langle A_\mathrm{POPC}\rangle \approx 0.61~\mathrm{nm}^2$ may
be due to differences in trans-gauche conformational changes.

As seen from Table~\ref{tabArea}, the presence of alcohol has a small but
non-vanishing effect on the area per lipid. The number of water molecules per
lipid molecule plays only a minor role as was verified by an additional
simulation of DPPC with ethanol and a reduced amount of water. Interestingly,
ethanol and methanol have almost the same effect on the area per lipid.

\begin{table}[b]
\begin{tabular}{@{}l|c|c|c|c@{}}
\hline \hline
System &    $d_1~[\mathrm{nm}]$ & $V_1~[\mathrm{nm^3}]$ & 
     $d_2~[\mathrm{nm}]$ & $V_2~[\mathrm{nm^3}]$ \\
\hline
DPPC (pure)  & $2.02\pm0.05$ &$1.32\pm0.03$ & $2.02\pm0.05$ &$1.32\pm0.03$\\
with ethanol & $1.84\pm0.02$ &$1.28\pm0.02$ & $1.90\pm0.02$ &$1.33\pm0.02$\\
with methanol & $1.93\pm0.06$ &$1.34\pm0.04$ & $1.95\pm0.06$ &$1.35\pm0.04$\\ 
\hline
POPC (pure)  & $1.96\pm0.04$ &$1.33\pm0.03$ & $1.96\pm0.04$ &$1.33\pm0.03$\\
with ethanol & $1.87\pm0.02$ &$1.31\pm0.02$ & $1.93\pm0.03$ &$1.35\pm0.02$\\
with methanol & $1.94\pm0.02$ &$1.35\pm0.01$ & $1.96\pm0.02$ &$1.36\pm0.01$\\
\hline \hline
\end{tabular}
\caption{The thickness $d$ of a leaflet (the bilayer thickness is twice that
value)
and the corresponding
volume per lipid using the two definitions given in Eq.~(\protect\ref{eq:vols}).
The error estimate for $d$ has been computed by cutting the analysis part
of the trajectory in two parts and by applying 
Eq.~(\protect\ref{eq:vols}) separately to both of them.
}
\label{tabVolu}
\end{table}

While the definition of the area per lipid is straightforward, the same
is not true for the volume occupied by a lipid.
The precise definition of the volume $V$ (or the
thickness $d$) of a membrane is non-trivial as discussed
at length by \citet{armen98}.
Here, we chose an operational definition based on
local mass density. Other definitions,
e.\,g., employing the electron density, are equally possible.

Below, we give the two definitions we used to compute the thickness.
If $\rho_{\text{lipid}}$, $\rho_{\text{water}}$ and
$\rho_{\text{alcohol}}$ are the mass densities of the three components,
the effective thickness of a single leaflet can be defined by
\begin{subequations}
\begin{gather}
d_1 = \frac{1}{2} \int \frac{\rho_{\text{lipid}}}{\rho_{\text{lipid}}
  + \rho_{\text{water}} + \rho_{\text{alcohol}} } \mathrm{d}z \;,\\
d_2 = \frac{1}{2} \int \frac{\rho_{\text{lipid}}}{\rho_{\text{lipid}}
  + \rho_{\text{water}} } 
\mathrm{d}z\;.
\end{gather}
\label{eq:vols}
\end{subequations}
These two definitions differ in their treatment of the alcohol volume
fraction and give the same thickness for pure lipid
bilayers. After defining the thickness, the volume is
simply $V=d \langle A \rangle$ with $\langle A \rangle$ being the average area per lipid. The results using both
of the above definitions are summarised
in Table~\ref{tabVolu}.

The thicknesses we obtained for pure POPC agree very well with recent x-ray
diffraction studies of \citet{Vogel:2000a} and computer simulation
studies of \citet{gullingsrud04} who obtained
$3.9~\mathrm{nm}$ and $3.92~\mathrm{nm}$, respectively, for the total bilayer
thickness $2\,d$. For DPPC the thickness and volume are a few percent larger
than the experimental results~\cite{Nag00}. Using the electron density to define
the thickness would have led to similar results, see
Sec.~\ref{secElectronDensity}.

A comparison of Tables~\ref{tabArea} and~\ref{tabVolu} shows that
the addition of ethanol or methanol to a bilayer expands its surface slightly
while the thickness decreases such that the volume per lipid does not change
significantly. This is as assumed since the main effect of the addition of
alcohol is a reduction of the surface tension of the water phase.
This is in agreement with observations from
a DPPC--halothane system~\cite{tu98}. 
We will return to this issue in Sec.~\ref{secDiscussion}.

\begin{figure}[b]
\centering
\includegraphics[height=3.8cm]{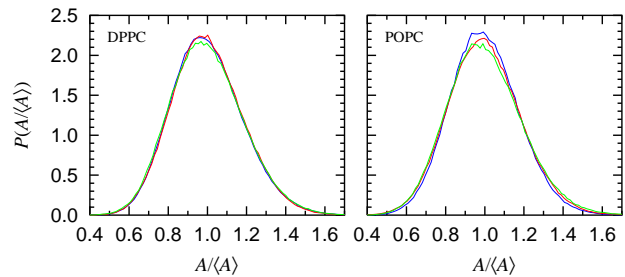}
\caption{Distribution of the individual areas of the lipids as determined by
two-dimensional Voronoi tessellation for DPPC (left) and POPC (right).}
\label{figAreaVoronoi}
\end{figure}

\begin{figure}[t]
\centering
\includegraphics[height=3.8cm]{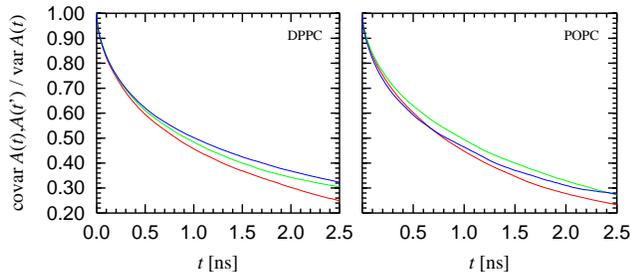}
\caption{Autocorrelation function for the individual areas of the lipids as 
determined by two-dimensional Voronoi tessellation for DPPC (left) 
and POPC (right).}
\label{figAreaCorrelation}
\end{figure}

\begin{figure}[b]
\centering
\includegraphics[height=3.8cm]{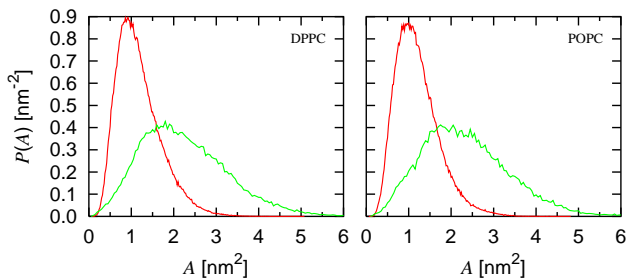}
\caption{Distribution of the Voronoi areas of the alcohols with
DPPC (left) and POPC (right).
Only alcohol molecules located close to the bilayer interface are included
in the analysis.}
\label{figVoronoiAlcohol}
\end{figure}

To complement the average area per lipid $\langle A\rangle$ measurements,
we have also computed the area probability distribution $P(A)$ by
Voronoi tessellation. 
By definition, Voronoi tessellation measures the area that is closer to
a given molecule than to any other one. The Voronoi approach does not uniquely
specify which point should be used to represent the entire molecule. 
We used the centre-of-mass position of each lipid, 
projected onto the $x y$-plane. Other choices are also possible, such as the
position of the \textit{sn}\,--\,3 carbon which gives a better indication of the backbone
of the lipid whereas the centre-of-mass describes the entire lipid.

The resulting distributions $P(A)$ are shown in Fig.~\ref{figAreaVoronoi}. 
The mean of that distribution is, by construction, identical to the average
area per lipid as shown in Table~\ref{tabArea}, and thus does not contain any
additional information. For this reason, not the plain distribution $P(A)$ but
rather the re-scaled distribution $P(A/\langle A\rangle)$ is shown in
Fig.~\ref{figAreaVoronoi}. Plotting the result in this way shows that alcohol
does not influence the Voronoi distribution in any way that would not be
captured already by the average area per lipid.

It is also possible to compute the autocorrelation
time of the Voronoi areas. This time gives an indication of how quickly
the geometry of the bilayer changes locally whereas the fluctuations in the
size of the simulation box seen in
Fig.~\ref{figAreaPerLipid} are related to global changes of the geometry. The
results are shown in Fig.~\ref{figAreaCorrelation}. The faster decay 
for the systems with ethanol suggests that the bilayer might
become more fluid but care should be taken in drawing conclusions as
the differences between the curves are rather small.

Next, we perform similar Voronoi tessellation for the alcohol molecules inside
the bilayer interface region. The precise definition of that region turned
out not to be critical, and we included all alcohol molecules within the
range $0.7~\mathrm{nm}<z<1.7~\mathrm{nm}$ from the centre of the bilayer. (Our
choice for this range is motivated by the results to be discussed later in
Sec.~\ref{secMassDensity}.) The variable number of molecules forbids a proper
calculation of the correlation time for the areas assigned to each alcohol,
though, and thus we present only the distribution $P(A)$ in
Fig.~\ref{figVoronoiAlcohol}. Since there are fewer methanol molecules close to
the bilayer than there are ethanols (see Fig.~\ref{figDensity1}),
the average area per methanol is larger than the average area per ethanol.

\subsection{Mass density}
\label{secMassDensity}

\begin{figure}[t]
\centering
\includegraphics[height=3.8cm]{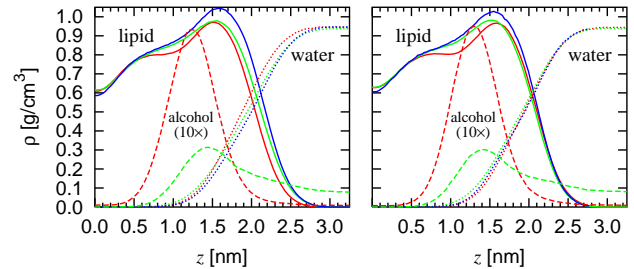}
\caption{Mass density profiles across the bilayer for DPPC (left) and POPC
(right). The density of the alcohol has been scaled by a factor of $10$, and
the colour code is the same as above.}
\label{figDensity1}
\end{figure}

The mass density profiles across the bilayer are shown in Fig.~\ref{figDensity1}.
For each analysed simulation frame, the system was first translated such
that the centre of the bilayer was located at $z=0$. Particles with $z<0$ were
mirrored to $z>0$ to reduce statistical error. The masses of the hydrogen
atoms were accounted for in the calculation.
Due to the low density of alcohol, its curve is scaled by a factor of $10$ in the
figure.

\begin{figure}[b]
\centering
\includegraphics[height=3.8cm]{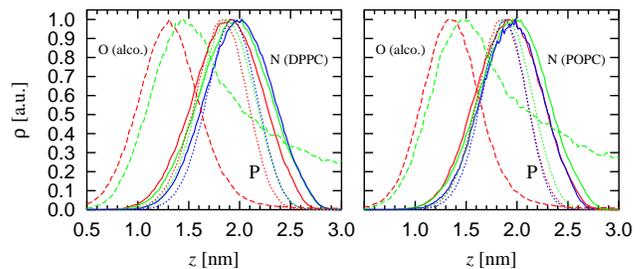}
\caption{Mass density profiles across the bilayer. The densities given
are the O(xygen) of alcohol as well as N(itrogen) and P(hosphorus)
of the lipid, and scaled to give a maximum of unity. Left for DPPC,
right for POPC.}
\label{figDensity2}
\end{figure}

\begin{figure}[t]
\centering
\includegraphics[height=3.8cm]{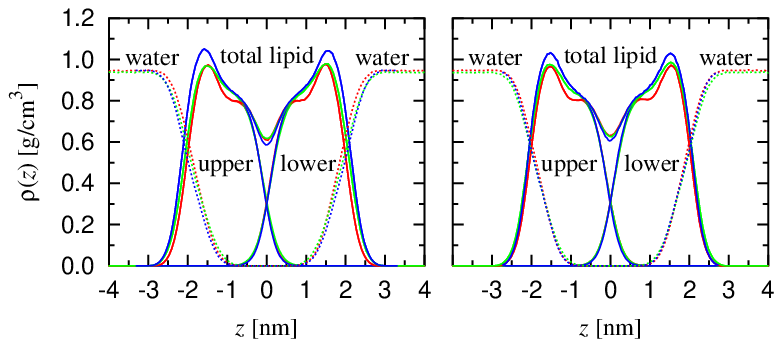}
\caption{Density profile across the whole bilayer. The lipid component is divided
into the contributions from the two separate leaflets, providing a 
measure of interdigitation. The alcohol component has been suppressed in
the figure. Left for DPPC, right for POPC.}
\label{figDensity3}
\end{figure}

Additional information can be gained by considering separately the two charged
groups in the lipid headgroups, namely the phosphate (P) and the choline (N)
group (cf. Fig.~\ref{figChemie}). In addition to this, the oxygen atom of the
alcohols is included in Fig.~\ref{figDensity2}. We could not find a direct
comparison for the mass density but the observations from the computer
simulations of \citet{Feller:2002} are consistent with our results for the mass
density. Figure~\ref{figDensity2} shows that the alcohol molecules have a
strong tendency to accumulate below the bilayer--water interface layer
(approximately given by the location of the phosphate and choline groups), and
that this tendency is stronger for ethanol than for methanol. We will return to
this issue in the partitioning analysis in Sec.~\ref{secPartitioning} and the
membrane penetration analysis in Sec.~\ref{secCrossings}.

The density of the lipid is decreased near the centre of the bilayer.
This phenomenon is known as lipid trough and means that the two leaflets are
repelling each other. Still, the tails of the lipids from one leaflet are able 
to penetrate into the other leaflet; this is known as 
interdigitation~\cite{Loebbcke:1995}.
To analyse this, in Fig.~\ref{figDensity3} we plot the density throughout
the whole bilayer, i.\,e., the positions of all atoms are not folded into
a single leaflet. We show separately the density of the lipids belonging to
the upper and the lower leaflet of the bilayer. It is easily seen that the
tails of the lipids can penetrate up to $0.5~\mathrm{nm}$ into the other
leaflet, and the degree of interdigitation is largely independent of the presence of 
alcohol.

\subsection{Electron density}
\label{secElectronDensity}

Electron densities provide information about the structure of
bilayers along the normal to the bilayer plane similar to
the mass densities.
Experimentally, x-ray diffraction provides a means to
access this quantity, the measurements yielding information of the
total electron density profile.

\begin{figure}
\centering
\includegraphics[height=3.8cm]{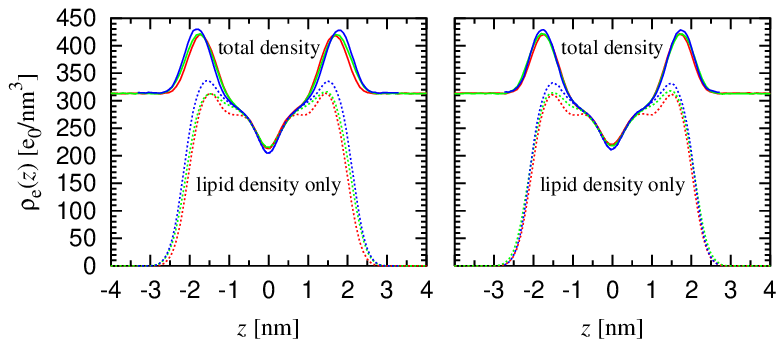}
\caption{Electron density profiles in the studied systems with
DPPC (left) and POPC (right).}
\label{figEleDensity}
\end{figure}

Figure~\ref{figEleDensity} shows the total electron densities in different
cases. For the pure lipid bilayers, the curves show the typical behaviour,
namely a maximum associated with the electron dense areas in the headgroup,
i.\,e., the phosphate groups, and the minimum at the bilayer centre -- the
so-called methyl trough~\cite{Nag00}. Experimentally, an electron density
profile contains different information than a mass density profile since
the chemical composition at depth $z$ is not known directly. (For computer
simulations this problem does not exist.)

For the pure systems our results agree well with
experiments~\cite{nagle96}. The only x-ray diffraction study in a
related system containing ethanol or methanol that we are aware of
is that of \citet{adachi}. Unfortunately a direct comparison
is not meaningful since that study was done in the gel phase using
multilamellar vesicles as compared to the fluid-like phase and
planar bilayer system studied here.

\subsection{Hydrogen bonding of alcohol to lipids}
\label{secBinding}

\begin{figure}
\centering
\includegraphics[width=7cm]{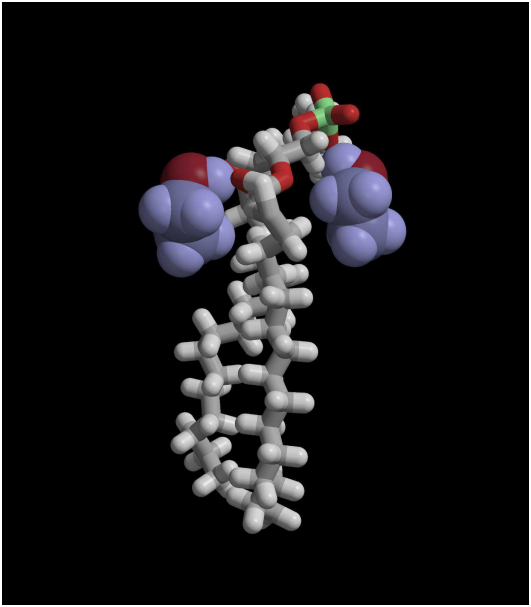}
\caption{A DPPC molecule together with two ethanol molecules. The ethanols are
located close to the ester oxygen. The DPPC molecule is drawn as rods whereas
the ethanols are drawn in a spacefilling representation. To aid the eye, the
ethanols are coloured blue-ish.}
\label{figZeichnung}
\end{figure}

As was shown in Sec.~\ref{secMassDensity}, the alcohol molecules have a
tendency to be located inside the bilayer, and this tendency is stronger for
ethanol than for methanol. The alcohol molecules are not located directly at
the water--membrane interface but rather further inside the bilayer. For the
simulations with ethanol, a direct visual inspection of the atom positions
shows that ethanol molecules are located close to the ester oxygens of the
lipids, see Fig.~\ref{figZeichnung}.

This visual conclusion is confirmed by a hydrogen bonding
analysis. In such an analysis, possible donors and acceptors are identified by
their chemical properties, and a hydrogen bond is then assumed to exist
whenever two such atoms and an additional hydrogen atom fulfil certain
geometric conditions. (The distance between a hydrogen and an acceptor has to be
smaller than $0.25~\mathrm{nm}$, and the angle between acceptor, hydrogen and 
donor has to be smaller than $60$ degrees.)

\begin{table}[b]
\begin{tabular}{l|rr}
\hline \hline
& DPPC & POPC \\
\hline
bound ethanols & $72.9$ & $71.6$ \\
bound lipids  & $59.7$ & $59.2$ \\
hydrogen bonds & $74.1$ & $72.8$ \\
lifetime [ns] & $1.20$ & $1.15$ \\
\hline \hline
\end{tabular}
\caption{Results of the hydrogen bonding analysis for the DPPC and POPC bilayers
with ethanol. (The systems consist of 128 lipid molecules
and 90 alcohol molecules.) No hydrogen bonds 
between methanol and lipids were found.}
\label{tabHydrogen}
\end{table}

The hydrogen bonding analysis shows that the majority of the ethanols are
involved in hydrogen bonds with lipids whereas not a single hydrogen bond
between a methanol and a lipid molecule was found in our simulations. The
results are summarised in Table~\ref{tabHydrogen}. Many lipids are involved in
more than one hydrogen bond which comes as no surprise since they possess an
ester oxygen in each of their two chains. Comparison with the lifetime data
for ethanol in Table~\ref{tabHydrogen} with NMR experiments~\cite{holte97:a}
shows excellent agreement. In their experiments Holte and Gawrisch reported
the lifetimes to be around $1~\mathrm{ns}$ while we obtained 
$1.20~\mathrm{ns}$ for
the ethanol lipid hydrogen bonds. We are not aware of any such experiments for
methanol.

The number of alcohol molecules involved in hydrogen bonds is best compared
against the total number of alcohol molecules located inside the bilayer. The
latter number is relatively ill-defined but from Fig.~\ref{figDensity1} one can
easily compute that for ethanol-containing systems only of the order of
$10$ ethanol molecules out of the approximately $70$ inside the bilayer are
not involved in hydrogen bonds. For comparison, in the methanol systems there
are of the order of $20$ methanol molecules inside the bilayer and none of 
them is involved in hydrogen bonds (but cf. Sec.~\ref{secCrossings}). 
These numbers
show that there indeed is a significant difference between ethanol and
methanol. 

Hydrogen bonding analysis offers a well-defined criterion to decide whether a
given lipid is interacting strongly with an alcohol molecule or not. This will
be used in the following sections to study separately the two lipid
populations, lipids bound to an alcohol and lipids not bound to an alcohol.

\subsection{Radial-distribution functions}

\enlargethispage{4mm}

\begin{figure}[b]
\centering
\includegraphics[height=3.8cm]{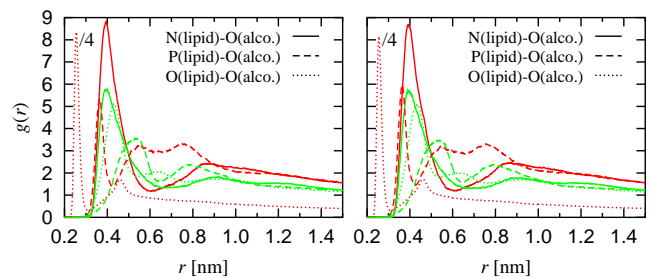}
\caption{Radial-distribution function between the oxygen of alcohol, on
the one hand, and 
the phosphorus and the nitrogen atoms in the
headgroup as well as the ester oxygens in the lipid tail, on the other hand.
The oxygen--oxygen curve has been scaled by a factor of $1/4$, i.\,e.,
in reality the RDF peaks at a value four times as large as displayed in
the figure.}
\label{figRdf1}
\end{figure}

\begin{figure*}
\centering
\includegraphics[height=3.8cm]{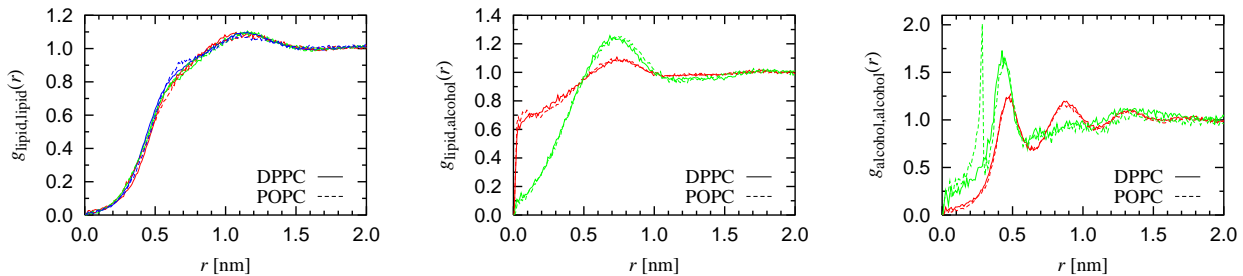}
\caption{Two-dimensional 
radial-distribution functions of the centre-of-mass
positions for lipid--lipid (left), lipid--alcohol (centre) and alcohol--alcohol
(right).}
\label{figRdf3}
\end{figure*}

In addition to the mass density profiles, valuable information may be gained
from radial-distribution functions (RDFs). The 
radial-distribution functions $g(r)$ give the probability
of finding two particles at a mutual distance $r$ once geometric and density 
factors have been scaled out.

Figure~\ref{figRdf1} shows
the RDFs between the oxygen of the alcohol and different charged groups inside
the lipid. Some of these groups were depicted already in the mass
density profile in Fig.~\ref{figDensity2} but whereas there only the vertical
distance between particles was considered, the RDF considers the real
three-dimensional distance between them.

While the mass density profile showed that on the average ethanol molecules
prefer to reside $0.5~\mathrm{nm}$ below the lipid headgroups, the RDF shows
that the three-dimensional preferred distance is only $0.38~\mathrm{nm}$. This
is no contradiction but
is easily understood by the observation (cf.
Fig.~\ref{figMassBinding} a bit further down) that lipid molecules without an attached ethanol
molecule are sticking out of the bilayer more than those with
an attached ethanol. This is captured only by the RDFs but not
by the mass density profile.

The radial-distribution functions for the different systems look quite similar
-- with one exception: The RDF between the alcohol and the ester group is peaked
at a much smaller distance for ethanol than it is for methanol. This is in
agreement with the results of the hydrogen bonding analysis.

By studying the mutual RDFs of the choline and\,/\,or phosphate groups, it is
possible to detect phase transitions of the bilayer. Within error margins, 
these RDFs are not dependent on the presence of alcohol, and for space reasons
we do not show them here as they are identical to the RDFs
published previously~\cite{patra:03aa}.

We have also studied two-dimensional radial-distribution functions
of entire molecules, i.\,e., the
molecules' centres-of-masses were projected onto the $x y$-plane and
radial-distribution functions were then computed. The results are shown in
Fig.~\ref{figRdf3}. The mutual radial-distribution functions of the lipids
exhibit a very soft core as lipids are able to wrap around each other. No
dependence on lipid type or presence of alcohol was observed. 

The RDF between alcohol and lipid is qualitatively different for ethanol
and methanol. For an ethanol, there is a large probability for it
to be at the same $x$-$y$ position as the (centre-of-mass of the) lipid. This
reflects the hydrogen bonding of ethanol close to the centre of the lipid. This
bonding is absent for methanol, and consequently then $g(r)\to 0$ for $r\to 0$.
No significant dependence on the kind of lipid is observed.

\begin{figure}[b]
\centering
\includegraphics[height=3.8cm]{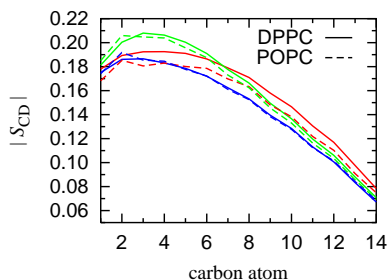}
\caption{Deuterium order parameters, computed from Eq.~(\ref{sOrderDefCD}), 
for DPPC (solid line) and POPC (dashed line).
For DPPC, the average over
the two tail chains is displayed while for POPC only the saturated 
\textit{sn}\,--\,1 chain is shown. For numbering of carbon atoms,
refer to Fig.~\ref{figChemie}.}
\label{figOrderParam}
\end{figure}

The alcohol--alcohol radial-distribution functions have a very different
character: For methanol with DPPC, the first peak is very distinct but the
correlations decay soon after. For methanol with POPC an additional peak is
observed. (This is the only curve with a significant difference between DPPC
and POPC. We cannot offer a convincing explanation for this.) For ethanol, the
behaviour shows almost quasi long-range order. The reason for this 
ordering is not clear and further experiments would be needed to study this in
detail.

\subsection{Order parameters}

Ordering of the lipid acyl chains is typically characterised
using the order parameter tensor
\begin{equation}
    S_{\alpha\beta} =
      \frac{1}{2}
      \left\langle
       3 \cos\theta_{\alpha}\,\cos\theta_{\beta} -
      \delta_{\alpha\beta}
      \right\rangle \;,
    \label{sOrderDef}
\end{equation}
where $\alpha,\beta = x,y,z$, and $\theta_{\alpha}$
is the angle between the $\alpha^{\text{th}}$
molecular axis and the bilayer normal ($z$-axis). The order
parameter is then computed separately for all carbons
along the acyl chain.
Since lipid bilayer systems possess symmetry with respect to
rotations around the $z$-axis, the relevant order parameter is the
diagonal element $S_{zz}$. To relate $S_{zz}$ to the experimentally
relevant deuterium order parameter
\begin{equation}
    S_{\mathrm{CD}} = \frac{2}{3} S_{xx} +
             \frac{1}{3} S_{yy} \;,
    \label{sOrderDefCD}
\end{equation}
we use the symmetry and write
$S_{xx} = S_{yy}$, and $S_{xx} + S_{yy} + S_{zz} = 0$.
Using these relations we have
$S_{\mathrm{CD}} = - S_{zz} / 2$.
To allow comparison with experimental data, we
present our results in terms of $\lvert S_{\mathrm{CD}}\rvert$.

\begin{figure}[b]
\centering
\includegraphics[height=3.8cm]{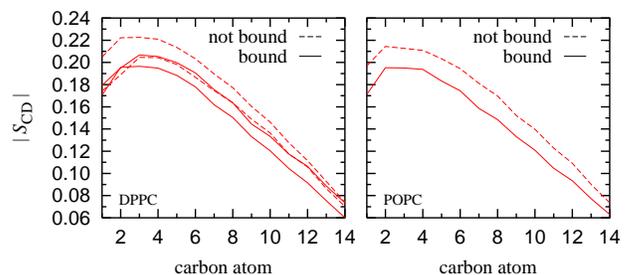}
\caption{Order parameter of DPPC (left) and POPC (right) in the presence of
ethanol. For DPPC, both the \textit{sn}\,--\,1 and \textit{sn}\,--\,2 chains are shown whereas for POPC
only the \textit{sn}\,--\,1 chain is depicted. The order parameter has been computed
separately for lipids that are bound to at least one alcohol molecule
(solid lines), and for
 lipids that are not bound to any alcohol (dashed lines).}
 \label{figOrderParam2}
\end{figure}

Since our simulations employ a united atom model,
no explicit information about the hydrogen positions is available and
they must be reconstructed assuming a perfect tetrahedral arrangement. The
results are shown in Fig.~\ref{figOrderParam}.
Reconstruction of the hydrogen
positions means that for the outermost carbon atoms of the tail no order
parameter can be constructed.
This also includes positions where a sequence of
carbon atoms connected by single bonds ends in a carbon atom having a
double bond. It is not a problem to compute the order parameter for a
chain of atoms connected by double bonds, there just is a problem connecting
such a chain to a chain of atoms connected by single bonds. For this reason
we show the order parameter for both chains of DPPC but for POPC
we restrict ourselves 
to the saturated \textit{sn}\,--\,1 chain, see Fig.~\ref{figChemie}.

\begin{figure}
\centering
\includegraphics[height=3.8cm]{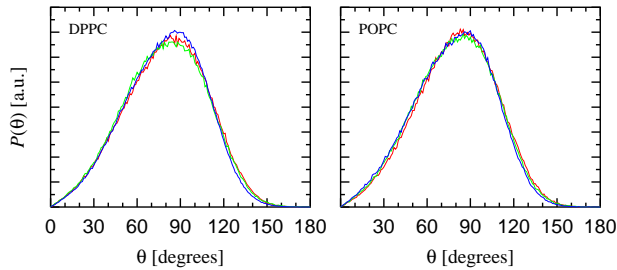}
\caption{Distribution of the angle $\theta$ between the P\,--\,N vector 
and the bilayer normal for
DPPC (left) and POPC (right). }
\label{figOrderHead}
\end{figure}

The results for the order parameter for all the cases are shown in
Fig.~\ref{figOrderParam}. We find that methanol increases ordering of the lipid
acyl chains close to the glycerol group, while the effect of ethanol is
strongest below the glycerol group, around the centre of the hydrocarbon tails.
These results are fully consistent with the mass density profiles in
Fig.~\ref{figDensity1}.

\begin{figure}[b]
\centering
\includegraphics[height=3.8cm]{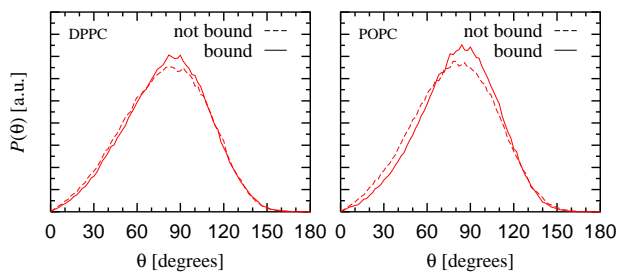}
\caption{Distribution of the angle between the P\,--\,N vector and the
bilayer normal for
DPPC (left) and POPC (right) in the presence of ethanol. The distribution is
separated into lipids that are hydrogen bonded respectively not hydrogen bonded
to an ethanol molecule.}
\label{figOrderHead2}
\end{figure}

As discussed in the introduction, there is, to our knowledge, only one other,
related membrane--alcohol simulation study~\cite{Feller:2002}. In that
the temperature was 40 degrees lower. Although direct comparison is not
possible, the order parameters in that study and ours are in qualitative
agreement. It is noticeable that the effect of alcohols on the average order
parameter is small -- this is consistent with the fact
that the volume occupied by a lipid changes only slightly.

Additional insight can be gained by combining the order parameters with the
hydrogen bonding analysis from Sec.~\ref{secBinding}. This combination allows
the computation of the order parameter depending on whether the lipid forms a
hydrogen bond with an alcohol molecule or not. The result in
Fig.~\ref{figOrderParam2} shows that binding of ethanol slightly decreases
the order of the tails.

In addition to the deuterium order parameter for the acyl chains, it is possible to
study the ordering of headgroups in a similar fashion. To do that, we have
chosen the angle of the P\,--\,N vector with respect to the bilayer interface
plane and have computed its distribution. The result is shown in
Fig.~\ref{figOrderHead}. The P\,--\,N vector has a significantly higher
tendency of being in the bilayer plane ($\theta=90\,^{\circ}$) than of
sticking out of it. The computed distribution is only slightly dependent on
the presence of alcohol.

Again, we can obtain additional information for the ethanol systems if the
angular distribution is separated into the distributions of lipids that are
hydrogen bonded to an alcohol, and those that are not. The result is shown in
Fig.~\ref{figOrderHead2}. For DPPC the angular distribution is
not influenced at all by hydrogen bonding whereas for POPC the influence is
small.

\begin{figure}
\centering
\includegraphics[height=3.8cm]{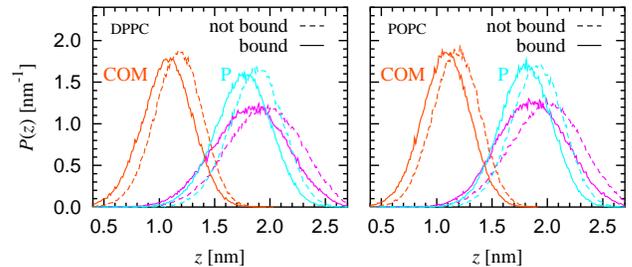}
\caption{Distribution of the position of the phosphate group (P), the
choline group (N) and the centre-of-mass of the entire lipid (COM),
left for DPPC, right for POPC.
Results are shown as solid (dashed) lines
for lipid molecules bound (unbound) to an ethanol molecule. Please note that in this figure the
colours do not mark the kind of alcohol present.}
\label{figMassBinding}
\end{figure}

To better explain the results presented in this section, we return to
the mass density profiles. We analysed the positions of the phosphate and the
choline groups in the heads of the lipids as well as the lipids' centre-of-mass
positions, separated into lipids that are hydrogen bonded or not hydrogen bonded to an ethanol.
The data in Fig.~\ref{figMassBinding} shows that lipid molecules are
shifted towards the centre of the bilayer by approximately $0.2~\mathrm{nm}$
if they are bonded to an ethanol molecule. This is in agreement with the reduced
order parameter as that is normally associated with a thinner bilayer. The
vertical distance between the choline and the phosphate group remains
unchanged, in agreement with the distribution of the P\,--\,N angle.

\subsection{Orientation of the water dipole}

\begin{figure}
\centering
\includegraphics[height=3.8cm]{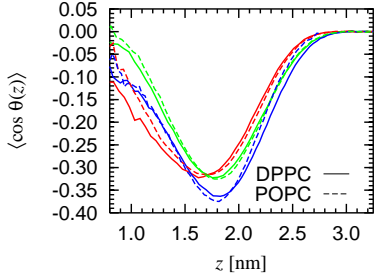}
\caption{Water orientation, as described by the mean cosine of the angle of the
water dipole moment with respect to the bilayer normal.}
\label{waterDipole}
\end{figure}

Ordering of the water dipole in the vicinity of the bilayer--water
interface is described by calculating the time averaged
projection of the water dipole unit vector $\vec{\mu}(z)$
onto the interfacial normal $\vec{n}$,
\begin{equation}
P(z) = \langle \vec{\mu}(z) \cdot \vec{n} \rangle
  = \langle \cos \theta(z) \rangle \;,
\end{equation}
where $z$ is the $z$-component of the centre-of-mass of the
water molecule and vector $\vec{n}$ points away from the
bilayer centre along the $z$-coordinate.

The data for all the studied cases are shown in Fig.~\ref{waterDipole}. For
pure bilayers the results are in agreement with previous studies, e.\,g.
\cite{patra:03b}. When either methanol or ethanol is added, the water dipole
becomes less oriented, i.\,e., the addition of alcohol slightly reduces the
amount of ordering. For pure bilayers, and for bilayers with added methanol,
the minimum remains at the same distance, at about $1.8~\mathrm{nm}$ from the 
bilayer centre. For added ethanol the minimum shifts to a smaller distance,
to about $1.6~\mathrm{nm}$. This is a reflection of the fact that ethanol 
leads to a slightly larger area per lipid and thus a thinner bilayer (see
Sec.~\ref{secDimensions}).

\subsection{Electrostatic potential}

To obtain the electrostatic potential across the bilayer the average charge
density profile was first computed such that the centre of the bilayer was
separately aligned to $z=0$ for each simulation frame. Then, the electrostatic
potential was determined by integrating the charge density twice with the
initial condition $V(z = 0) = 0$.

\begin{figure}[t]
\centering
\includegraphics[height=4.75cm]{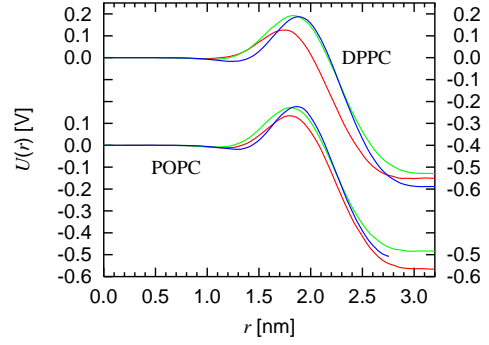}
\caption{Electrostatic potential through the bilayer for DPPC (top) and POPC
(bottom).}
\label{figPotential}
\end{figure}

The electrostatic potentials for all studied cases are shown in
Fig.~\ref{figPotential}. For pure DPPC the electrostatic potential was
determined to be $-589~\mathrm{mV}$ in agreement with previous
studies~\cite{tieleman:96a,patra:03b}. For pure POPC we obtain
$-507~\mathrm{mV}$.

The addition of alcohol leads only to small changes in the electrostatic
potential. This is what is expected from the results presented so far. The only
way in which the membrane potential could be changed significantly would be by
re-arrangement of the P\,--\,N angle of the headgroup, resulting in a change of
the dipole moment of the lipid headgroup. No such re-arrangement was observed, 
cf.~Fig.~\ref{figOrderHead}.

On a superficial level this may seem to be in contradiction to some previous
suggestions that the narcotic effects of alcohols are mainly due to a change of
the electrostatic potential. We would like to point out, however, that even
though the direct effect of alcohol to the potential is small, this does not
exclude secondary effects which may lead to a significant change in the
electrostatic potential. We return to this issue in Sec.~\ref{secDiscussion}.

\subsection{Partitioning coefficients}
\label{secPartitioning}

To measure whether alcohols prefer to stay close to the membrane or rather to
be in the water phase, we computed the partitioning coefficients. They
provide a thermodynamic quantity characterising a molecule's tendency
to choose its environment.

The partitioning coefficient $K_{\text{p}}$ is defined as
\begin{equation}
	K_{\text{p}} = X_{\text{bilayer}} / X_{\text{water}}\;,
	\label{partcoeff}
\end{equation}
where $X_{\text{bilayer}}$ and $X_{\text{water}}$ are the moles of solute per
kg of solvent~\cite{westh:01a}. Equation~(\ref{partcoeff}) can be reformulated 
in terms of the molar concentrations $n$ and the molar masses $m$ as
\begin{equation}
    K_{\text{p}} = \frac{( 1 - n^{\text{water}}_{\text{alcohol}})
    	n^{\text{bilayer}}_{\text{alcohol}}
    m_{\text{water}}}{
        (1 - n^{\text{bilayer}}_{\text{alcohol}})
	    n^{\text{water}}_{\text{alcohol}}
        m_{\text{lipid}}
    } \;.
\end{equation}
Here, $n^{\text{a}}_{\text{b}}$ means the molar concentration
of component $b$ in the phase $a$.

\begin{table}[t]
\mbox{}\hspace{-5mm}\begin{tabular}{l|c|c|c}
 \hline \hline
system & $n^{\text{water}}_{\text{ethanol}}$
    & $n^{\text{lipid}}_{\text{ethanol}}$ & $K_{\text{p}}$ \\[0.8mm]
 \hline
DPPC--ethanol & $(3.98\pm0.06)\cdot 10^{-4}$ 
    & $0.403\pm0.001$ & $41.6\pm0.6$ \\
POPC--ethanol & $(3.34\pm0.04)\cdot 10^{-4}$ 
    & $0.405\pm0.001$ & $48.3\pm0.9$ \\ \hline
DPPC--methanol & $(4.53\pm0.01)\cdot 10^{-3}$ 
    & $0.278\pm0.001$ & $2.07\pm0.02$ \\
POPC--methanol & $(4.72\pm0.02)\cdot 10^{-3}$ 
    & $0.272\pm0.001$ & $1.87\pm0.02$ \\
\hline \hline
\end{tabular}
\caption{Results of the partitioning analysis.}
\label{tabPartitioning}
\end{table}

One might be tempted to count the number of alcohol molecules in the lipid
phase and in the water phase using some functional definition of where the
lipid bilayer interface is located. This approach, however, gives incorrect
results as the relevant quantity is not the number of alcohol molecules 
\emph{inside} the membrane but rather the number of alcohol molecules
\emph{influenced} by the membrane.

Starting point are the mass densities $\rho_{\text{water}}$ and
$\rho_{\text{alcohol}}$ far from the bilayer in the water phase. These are
available from Fig.~\ref{figDensity1}, and knowledge of the molar masses gives
$n_{\text{water}}$. Multiplying $n_{\text{water}}$ by the
total number of water molecules in the simulation box gives the number of
alcohol molecules that would be there if there was no lipid bilayer. The
remaining alcohol molecules thus must be ``brought in'' by the lipid bilayer.

Our results are summarised in Table~\ref{tabPartitioning}. Although experimental
measurements of partitioning coefficients have been 
performed~\cite{Rowe:1998,Westh+Trandum:1999,trandum00}, we were not able to
find experimental data to make exact quantitative comparisons. However, the
experimental data for DMPC~\cite{trandum00} with ethanol is in qualitative
agreement with our results. They are also in agreement with the general trends
that $K_{\text{p}}$ increases as the lipid tails or the alcohols
become longer. As we have studied the system at single temperature we
cannot comment on the temperature dependence of the partitioning
coefficients.

\subsection{Crossing events}
\label{secCrossings}

\begin{figure}[b]
\centering
\includegraphics[height=3.8cm]{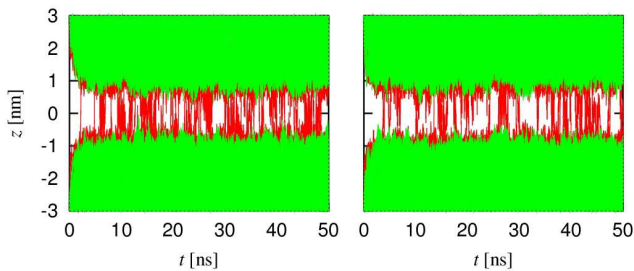}
\caption{$z$-positions of all alcohol molecules as a function of time for DPPC
(left) and POPC (right). Ethanol (red) is able to penetrate into the bilayer
(located at $z=0$) much better than methanol (green). Crossing events of ethanol
are seen while they are completely absent for methanol.}
\label{figAlleZ}
\end{figure}

\begin{figure*}
\centering
\includegraphics[height=3.8cm]{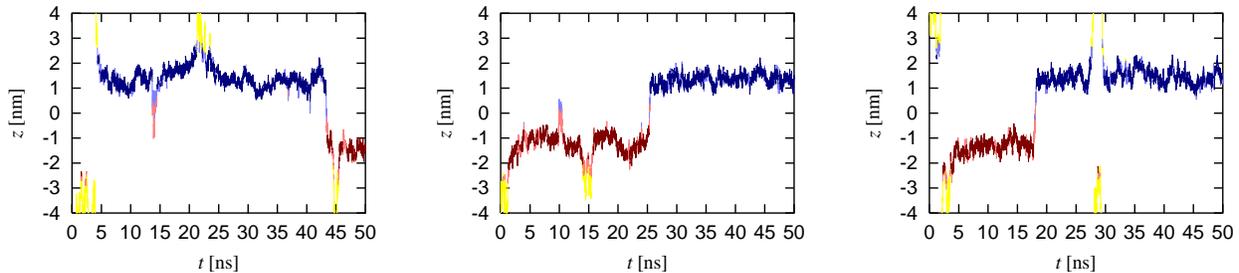}
\caption{$z$-component of the centre-of-mass position of a few tagged ethanol
molecule in a DPPC bilayer system. Each figure depicts the trajectory of
a different molecule. We have chosen three out of the $90$ molecules to
give a demonstration of the possible behaviour. The colours red and blue 
mark in which
leaflet of the bilayer the alcohol is located. If the colour is dark red or dark
blue, the
alcohol is hydrogen bonded to a lipid, otherwise the colour is light red or
light blue. If the alcohol is in the water phase, the trajectory is shown
in yellow.}
\label{figTrajExamples}
\end{figure*}

Next, we analyse the penetration of alcohol through the membrane. 
A quick overview can be obtained by
plotting the $z$-component of the positions of the alcohol molecules. (The
positions of all atoms have to be translated for every simulation frame such
that the centre of the bilayer does not move). The result is shown in
Fig.~\ref{figAlleZ}. The density (in space and time) of alcohol molecules is
large in most parts of the diagram such that it is difficult to visually
identify individual trajectories.

It is easily seen from Fig.~\ref{figAlleZ} that the density of alcohol molecules
is reduced in the centre of the bilayer. In addition, the gap in the centre of
the bilayer is smaller for ethanol than it is for methanol. This is in
agreement with the mass density profiles presented in Sec.~\ref{secMassDensity}.
It is also evident that there is a significant number of events where an
ethanol molecule is crossing from one leaflet to the other while no such
events are seen for methanol molecules. 
(Crossing events cannot be inferred from
the mass density since these events happen so fast that the resulting mass
density of alcohol in the centre of the bilayer is negligible.)

In the
simulations, it is directly known which atoms form the lipid molecules of the
upper leaflet of the bilayer, and which atoms form the lower leaflet, and
which atoms belong to water molecules. The atom nearest to the alcohol molecule
then determines in which of the three phases a given alcohol molecule is
located at any given moment. In addition, it is also relevant whether an
alcohol molecule is hydrogen bonded to some lipid molecule, cf.
Sec.~\ref{secBinding}. 

When all these pieces of information are combined, one arrives at data as shown
in Fig.~\ref{figTrajExamples} in which we depict a few selected ethanol
molecules within a DPPC bilayer. It is seen that, while the alcohol is inside the
bilayer, it is hydrogen bonded most of the time. The bonding does not persist
for the entire duration of a simulation but there are short breaks in between.
This is in agreement with the hydrogen bond lifetime of order $1~\mathrm{ns}$
in Table~\ref{tabHydrogen} while ethanol molecules can stay inside the bilayer
much longer than this. Whenever the hydrogen bond is broken, it can either be
re-formed shortly afterwards, or the alcohol molecule can try to move to some
other place. From the figure it is seen that an alcohol molecule can move to
the opposite leaflet of the bilayer but that not all such attempts are
successful, i.\,e., the alcohol molecule may be reflected back.

\begin{table}[b]
\begin{tabular}{l|cc|cc}
\hline \hline
 & \multicolumn{2}{c}{successful} & \multicolumn{2}{c}{unsuccessful} \\
System & number & time [ps] & number & time [ps] \\
\hline
DPPC--ethanol & 30 & 325 & 123 & 245 \\
POPC--ethanol & 21 & 375 & 101 & 225 \\
\hline \hline
\end{tabular}
\caption{Number of successful and unsuccessful crossing events, respectively, 
within $40~\mathrm{ns}$ of trajectory. In addition, the mean time spent in
the crossing process is given.}
\label{tabCrossings}
\end{table}

Using the collected information, each alcohol molecule is at any given moment
in one of five different states (water phase, upper leaflet, upper leaflet
hydrogen-bonded, lower leaflet and lower leaflet hydrogen-bonded; we will
discuss the methanol-containing systems a bit further down). It is of little
interest if an alcohol molecule is ``scratching'' at the surface of the bilayer
-- rather it is important whether the alcohol molecule reaches the part of
the leaflet where it can form a
hydrogen bond.

\begin{figure}[b]
\centering
\includegraphics[width=7cm]{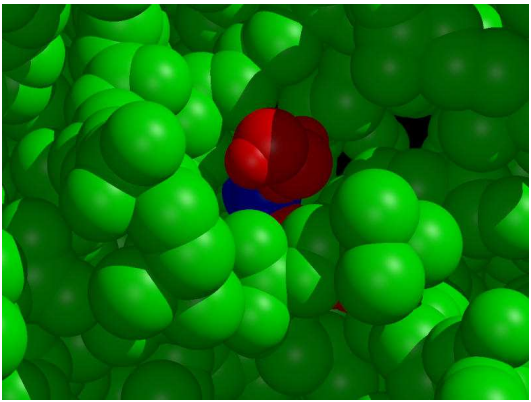}
\caption{View from the top onto part of a POPC bilayer with methanol. The lipids
are coloured green, methanol is coloured blue and a few selected water molecules
are shown in red. It is easily seen than the methanol is located in a cavity
together with a few water molecules.}
\label{figTopView}
\end{figure}

This immediately gives functional definitions for different kinetic events
that can be used for an automatic analysis. A successful crossing event from
the upper to the lower leaflet is for example given by a sequence ``upper leaflet
hydrogen-bonded $\to$ \{upper leaflet\} $\to$ \{lower leaflet\} $\to$ lower
leaflet hydrogen-bonded'' where the curly braces mean that this step may also be
skipped. Similarly, an unsuccessful crossing from top to bottom would be ``upper
leaflet hydrogen-bonded $\to$ \{upper leaflet\} $\to$ lower leaflet $\to$
\{upper leaflet\} $\to$ upper leaflet hydrogen-bonded''. Other criteria
are constructed similarly.

Table~\ref{tabCrossings} shows the results of the analysis for
crossings of ethanol molecules between the two leaflets. A simple calculation
shows that ethanol molecules are able to move from one leaflet to the other on
a time scale of $130~\mathrm{ns}$ for DPPC and $180~\mathrm{ns}$ for POPC.
The number of unsuccessful crossing attempts outnumbers the number
of successful attempts by a factor of $4$, thereby demonstrating that the
hydrophobic tails of the lipid pose a significant barrier to ethanol not only
from the outside of the leaflet but also from the inside. For methanol we did
not find any crossing events in our simulations, implying that the 
corresponding time scale must be at least of the order of microseconds. 

A closer study of the systems containing methanol is hindered by a
problem that is not obvious from any of the data presented so far. As
Fig.~\ref{figTopView} shows, methanol is virtually never really inside the
bilayer, i.\,e., located such that it no longer has direct contact with the bulk
water phase -- in all of our data for DPPC and POPC with methanol, we found only a
single methanol molecule that had actually lost contact with water.

\begin{figure}[b]
\centering
\includegraphics[height=3.8cm]{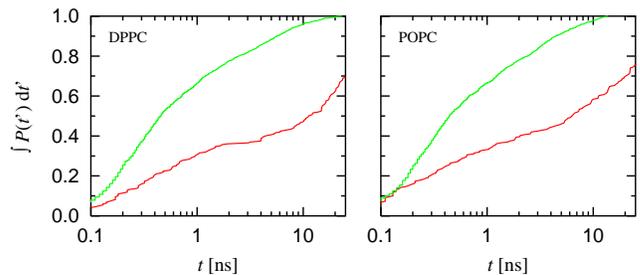}
\caption{Distribution $P(t)$ of the time $t$ for which an alcohol molecule
stays inside the membrane, left for DPPC, right for POPC. Due to limited
statistics, we cannot plot $P(t)$ directly but are limited to the cumulative
probability $\int P(t') d t'$.}
\label{figSurvivalTime}
\end{figure}

The above observation is also able to explain why no hydrogen bonds are formed
between methanol and the lipids: Water is simply too energetically favourable a
binding partner for methanol, or, in other words, the surface tension of water
is too high for methanol to leave the water phase. The concept of an alcohol
being inside the membrane thus does not apply -- topologically, methanols are
always located outside the membrane. Rather we need to introduce the concept
of a methanol being located in a sufficiently deep well. This can be quantified
by counting the number of atoms belonging to lipids within a certain distance
around some particular methanol molecule. This number will be much larger if the methanol
is inside such a well. (We use the criterion that the number of atoms 
belonging to lipid molecules within
$0.6~\mathrm{nm}$ is larger than $50$.) 

Using that functional definition, we are able to treat ethanol and methanol
containing systems on a
similar footage. While there are no crossing events for methanol, another
interesting question still arises, namely the dynamics of alcohol exchange
between the membrane and the water phase. Quantitatively, the interesting
quantity is the time $t$ between an alcohol molecule entering the membrane and
its subsequent leaving it again. Our results are shown in
Fig.~\ref{figSurvivalTime}. Since there are only of the order of $200$ ($1000$) events 
for ethanol (methanol), the statistics is insufficient to compute the
probability distribution $P(t)$. Rather, we present the cumulative probability
$\int P(t') d t'$, i.\,e., the probability that an alcohol stays inside the
membrane no longer than some time $t$, since this quantity can be computed
without binning the data point.
($P(t)$ follows, in principle, by differentiation of the
depicted curves.) It is seen from the figure that the dynamics is much faster
for methanol than for ethanol. This comes as no surprise since methanol is
not really inside the bilayer -- it does not need to cross the bilayer
interface but only needs to deform it (to create a well).

\section{Discussion}
\label{secDiscussion}

This study is, to the best of our knowledge, the first detailed computational 
study characterising the behaviour of lipid bilayers (POPC and DPPC) under the
influence of methanol and ethanol. The other existing molecular dynamics study
of ethanol and POPC~\cite{Feller:2002} concentrated on the comparisons with an
NMR study under different conditions (at 10 degrees Celsius close to the gel
phase and using an $NVT$ ensemble). To obtain detailed information about
alcohol--membrane interactions, it is thus important to study the biologically
important fluid phase.

Let us first discuss the area per lipid and bilayer thickness. The increase in
the area per lipid is larger for DPPC bilayers (about $7\,\%$ for ethanol and
$6\,\%$ for methanol) as compared to POPC systems ($5\,\%$ for ethanol and
$3\,\%$ for methanol). This compares well with the recent micropipette studies
of \citet{ly02} who used SOPC vesicles under slightly
different conditions ($20$ vol.-$\%$ ethanol at room temperature). They observed
$9\,\%$ increase in the area per lipid and $8\,\%$ decrease in the thickness
of the bilayer. Here, we obtained a decrease of $7\,\%$--$10\,\%$ (ethanol and
DPPC) and $1\,\%$--$4\,\%$ (ethanol POPC) in thickness depending on the
definition used, see Eq.~(\ref{eq:vols}). 

As a purely structural effect, it is clear that the membrane thus becomes more
permeable to small molecules due to its increased area per lipid. The
differences between DPPC and POPC are most likely due to the slightly longer
\textit{sn}\,--\,2 chain of POPC and the double bond in it. In addition to the effects
captured by the average area per lipid, steric constraints seem to make the POPC
bilayer less susceptible to penetration of small solutes.

Furthermore,
in a recent study \citet{chanturiya:1999} proposed
that the penetration of alcohols inside the bilayer, and their binding at it,
and the resulting decrease in bending rigidity is a feasible pathway for
promoting fusion of cells. Although it is not possible to probe this directly
by current computer resources, our observations support the
possibility of such a mechanism.

It has been suggested that the preferred location of ethanol close to the
membrane dehydrates it~\cite{Klemm:1990,Klemm:1998,holte97:a}. This
should show in the water dipole orientation data, Fig.~\ref{waterDipole}. The
relatively small changes in it, and in the
electrostatic potential across the membrane, suggest that indirect effects, such
as receptor blocking, may be more important in producing changes in these
quantities.

To characterise thermodynamic properties, we have measured the partitioning
coefficients. Our estimates of $K_\mathrm{p}$ for the different alcohol-lipid
combinations are given in Table~\ref{tabPartitioning}. Due to the lack of
experimental data, no quantitative comparison could be made but qualitatively
experiments and simulations are in agreement.

Short-chain alcohols have an amphiphilic character and it has been known for
long that addition of each new CH$_2$ group -- adding a CH$_2$ group on 
methanol gives ethanol and so on -- has a strong effect on the interactions
with membranes. This is indicated by the well known Traube's
rule~\cite{traube1891,adamson97} which states that the addition of a new 
CH$_2$ group leads to a decrease in surface tension. In other words,
short-chain alcohols have a strong effect on membrane properties and the
effect depends on both the length of the hydrophobic part of the
alcohol and on concentration. This has also been observed in recent
experiments~\cite{ly02,ly-pre}. 

Our data for methanol and ethanol supports these conclusions.
Methanol does not penetrate through the lipid tail region which is easily understood
by the hydrophobic nature of the lipid tails which are repelling methanol as it
is more polar than ethanol. As a second effect, methanol rarely reaches the tail
region as each methanol molecule moves together with a small cluster of water
molecules when it is trying to enter the membrane. This means that, on one hand,
methanol is pulled back into the water phase by this, and, on the other hand,
not a single small methanol molecule but a significantly larger dressed
particle, or a small cluster, would need to penetrate the membrane.

The analysis of crossing events, i.\,e., how often the molecules travel
through the membrane, showed that ethanol is able to penetrate the membrane
easily whereas for
methanol not a single crossing event was observed. This confirms the interpretation
given above.
It is difficult to compare these results
directly with experiments but the possibility of such crossing events has been
proposed on the basis of NMR studies~\cite{holte97:a}. The results presented
here are, to our knowledge, the first detailed analysis of crossing events.
Further experiments would be needed in order to better characterise the situation
as the system here is a simple model system and the relevance of these
results to biological systems, in particular yeasts, needs to be better
studied. The only such a study we were able to find uses NMR and \textit{Z.
mobilis}~\cite{Schobert:1996} but direct comparison is not possible due to the
different experimental setup.

In the introduction we briefly discussed general anaesthesia and
membrane--protein interactions induced by the addition of anaesthetics, such as
small alcohols. This was observed in a recent experiment~\cite{Brink:2004}
where the potassium channel KcsA was observed to dissociate due to the changes in
lateral membrane pressure induced by small alcohols. Here, we have
characterised simple membrane--alcohol systems. The detailed
characterisation presented here is essential for extensive simulational
studies of membrane--protein--anaesthetic systems. From our results it
is obvious that the changes in pure membranes are subtle but the effects of
those changes to, e.\,g., embedded proteins may be 
significant~\cite{cantor97,eckenhoff01,koubi01,tang:2002,Brink:2004}. This is
also supported by recent experiments using enflurane and DPPC~\cite{hauet03}.
Similar conclusions have been drawn by
\citet{tu98} for the interaction of halothane with bilayers.
As pointed out by Hauet~et~al., there are various intriguing questions
regarding small molecules and anaesthesia. These questions are related to 
interactions between membranes and small molecules and computer simulations 
give a direct access to study them.

\begin{acknowledgments}
We are grateful to Ole G. Mouritsen, Amy Rowat, 
Margie Longo, and John Crowe
for fruitful discussions.
This work has, in part, been supported the European Union through
Marie Curie fellowship program No.~HPMF--CT--2002--01794 (M.\,P.),
the Academy of Finland
through its Centre of Excellence Program (E.\,S., E.\,T. and I.\,V.), and
the Academy of Finland Grant Nos.~54113, 00119 (M.\,K.), 80246 (I.\,V.),
and 202598 (E.\,T.).
We would like to thank the Finnish IT Centre for Science
and the HorseShoe (DCSC) supercluster computing facility
at the University of Southern Denmark for computer resources.

\end{acknowledgments}

\end{document}